\documentclass{article} 
\usepackage{iclr2026_conference,times}

\usepackage{amsmath,amsfonts,bm}

\def\eqref#1{equation~\ref{#1}}

\def\1{\bm{1}}

\DeclareMathAlphabet{\mathsfit}{\encodingdefault}{\sfdefault}{m}{sl}
\SetMathAlphabet{\mathsfit}{bold}{\encodingdefault}{\sfdefault}{bx}{n}

\usepackage{hyperref}
\usepackage{url}

\usepackage{xcolor} 
\usepackage{ifthen}
\usepackage{soul} 
\usepackage{amssymb} 
\usepackage{adjustbox} 
\iclrfinalcopy

\newboolean{showcomments}
\setboolean{showcomments}{false} 
\ifthenelse{\boolean{showcomments}}
 { \newcommand{\mynote}[2]{
      \fbox{\bfseries\sffamily\scriptsize#1}
        {\small$\blacktriangleright${{{#2}\bf }}$\blacktriangleleft$}}}
        { \newcommand{\mynote}[2]{}}

\newcommand{\ak}[1]{\mynote{Ahmed}{\hl{#1}}}

\title{Impact of LLMs news Sentiment Analysis on Stock Price Movement Prediction}

\author{
{Walid Siala$^1$}, 
{Ahmed Khanfir$^{2,1}$} and 
{Mike Papadakis$^1$}\\
{$^1$ SnT, University of Luxembourg, Luxembourg}\\
{$^2$ RIADI, ENSI, University of Manouba, Tunisia}\\
\texttt{wsiala4@gmail.com},\\ 
\texttt{ahmed.khanfir@ensi-uma.tn} and \\
\texttt{michail.papadakis@uni.lu}
}

\begin{document}

\maketitle

\begin{abstract}
This paper addresses stock price movement prediction by leveraging LLM-based news sentiment analysis. 
Earlier works have largely focused on proposing and assessing sentiment analysis models and stock movement prediction methods, however, separately. 
Although promising results have been achieved, a clear and in-depth understanding of the benefit of the news sentiment to this task, as well as a comprehensive assessment of different architecture types in this context, is still lacking. Herein, we conduct an evaluation study that compares 3 different LLMs, namely, DeBERTa, RoBERTa and FinBERT, for sentiment-driven stock prediction. 
Our results suggest that DeBERTa outperforms the other two models with an accuracy of 75\% and that an ensemble model that combines the three models can increase the accuracy to about 80\%. Also, we see that sentiment news features can benefit (slightly) some stock market prediction models, i.e., LSTM-, PatchTST- and tPatchGNN-based classifiers and PatchTST- and TimesNet-based regression tasks models.
\end{abstract}

\section{Introduction}
\label{sec:introduction}

Stock price prediction has been an active research topic due to its importance in investment strategies and financial risk assessment. As the financial market fluctuates significantly due to multiple economic, technical, and political factors, the precise prediction of stock prices remains extremely challenging.  One research line has proposed to leverage news sentiment analysis, suggesting that sentiment-driven reactions are directly correlated with the variation of the stock price~\cite{baker2007investor,shiller2017}.

The introduction of large language models (LLMs) such as BERT~\cite{devlin2019bert}, revolutionized NLP, including text sentiment analysis, by facilitating deep contextual representation learning. Consequently, Domain-specialized variants, such as FinBERT \cite{araci2019finbert} have achieved state-of-the-art results on well-known financial sentiment datasets \cite{zhang2023domain} compared to traditional machine learning approaches~\cite{araci2019finbert,he2021deberta}.

Nonetheless, these pioneering works on LLM-based sentiment analysis for stock prediction are still in their infancy, necessitating a deeper and more extensive understanding of their capabilities and limitations. In fact,  (1) existing works tend to investigate one individual LLM independently or without comparing it to other models on the same labeled financial dataset; (2) although LLM sentiment outputs can be encoded under different representations, such as probability scores, and discrete labels, a study including these different representations and daily aggregation methods is missing; and (3) the fusion of sentiment analysis outputs with time-series stock data is underexplored, especially with recent state-of-the-art architectures such as patch-based transformers \cite{nie2023patchtst}, temporal graph neural networks \cite{li2023tpatchgnn}, and temporal convolution models \cite{wu2023timesnet}.

To fill these gaps, this paper presents an evaluation study that assesses the use of different LLM-based sentiment analysis approaches for stock prediction. In particular, it compares three LLM architectures, including FinBERT~\cite{araci2019finbert}, RoBERTa~\cite{liu2019roberta}, and DeBERTa~\cite{he2021deberta}, for financial sentiment classification.
The latter are then integrated with traditional and recent state-of-the-art time-series models for stock prediction, such as Long Short-Term Memory (LSTM) networks \cite{hochreiter1997long}, TimesNet \cite{wu2023timesnet}, PatchTST \cite{nie2023patchtst}, and tPatchGNN \cite{li2023tpatchgnn}.
The obtained results on well-known datasets demonstrate that the sentiment analysis models are complementary and can achieve about 80\% of accuracy when combined into an ensemble model, i.e. by training an SVM model on their outputs. 
The obtained news sentiment features can improve the stock price predictions, particularly in classification tasks, improving the accuracy of LSTM-, tPatchGNN- and PatchTST-based classifiers. It also improves the performance of regression models based on TimesNet and PatchTST architectures. 

\ak{talk about aggregation}

\ak{background and related work}

\section{Proposed Experimental Study}
\label{sec:setup}
\paragraph{Sentiment analysis models:}
We have included 6 sentiment analysis models: 3 well-established Transformers-based ones and three ensemble models that we trained using the combination output of the three former. 
The transformers-based models are:
FinBERT~\cite{araci2019finbert}, a domain-adapted variant of BERT~\cite{devlin2019bert} finetuned on large-scale financial corpora; 
RoBERTa~\cite{liu2019roberta} and DeBERTa~\cite{he2021deberta}, two general-purpose optimised versions of BERT, which perform efficiently on finance-specific tasks~\cite{sy2023fine}.
For a given input text, they output a sentiment class (negative, neutral or positive) and the prediction confidence score.  
They are deterministic by default, non-generative and thus less prone to hallucinations, enforcing the reproducibility of our work and reducing the impact of randomness on our results.
Using their predictions on a subset of SEntFiN~\cite{sinha2022sentfin10} dataset, We trained three ensemble models, 
via traditional machine learning algorithms: Random Forest (RF), Logistic Regression (LR), and Support Vector Machines (SVM).

\paragraph{Stock prediction methods:}
For the downstream task of stock movement prediction, we adopt four time-series forecasting methods representing fundamentally different architecture families. Long Short-Term Memory (LSTM) network \cite{hochreiter1997long} is included as a baseline for sequential modeling. PatchTST \cite{nie2023patchtst} and TimesNet \cite{wu2023timesnet} represent transformer-based architectures optimized for multivariate time-series analysis, with PatchTST utilizing a patching mechanism for temporal segmentation and TimesNet employing temporal variation modeling in two-dimensional space. tPatchGNN~\cite{li2023tpatchgnn} extends these capabilities through the integration of temporal patching with graph neural networks, enabling the capture of both temporal dynamics and relational dependencies between assets.

\paragraph{Datasets:}
The study examines five major publicly traded companies -- Microsoft (MSFT), Amazon (AMZN), Apple (AAPL), Netflix (NFLX), and Tesla (TSLA) -- over the period from March $10^{th}$ 2022 to April $2^{nd}$ 2025.
We collect the related daily stock data from Yahoo Finance
and the corresponding news (total of over 96000) via the AlphaVantage API. 
News are assigned to trading days according to their release timestamps, with weekend and holiday publications mapped to the next available trading day. Any missing values in the price series are handled through forward filling for continuity in time-series modeling. 
In order to evaluate the efficiency, complementarity of the Financial sentiment analysis models and to train the ensemble models, we have used a labelled financial news dataset; SEntFiN 1.0~\cite{sinha2022sentfin10}.
Unlike the Alphavantage dataset, SEntFiN provides news as text-sentiment pairs (over 10,700 news headlines), enabling model evaluation, but without any information about their issuing date, or related stock, which makes it unsuitable for the rest of the study. 

\paragraph{Proposed daily sentiment representations:}
Because multiple news items may be released on the same trading day, we apply aggregation procedures to group individual article-level sentiments into a single daily feature set. 
To do so, we compute the daily sum of sentiment scores to capture overall sentiment intensity, the minimum and maximum scores to reflect the most pessimistic and most optimistic signals of the day, and the majority-vote class to summarise the prevailing sentiment.
The resulting dataset contains both market-based variables (daily volume and closing price) and sentiment-based variables in each aggregated representation format. 

\paragraph{The prediction targets} are defined in two forms: price factor (regression task) and binary trend movement (classification task). Formally, let $P_t$ denote the closing price at trading day $t$, and let $\tilde{P}_t = P_{t+1}$ represent the next-day closing price aligned to time $t$. 
The factor (price ratio) is computed as $\text{Factor}_t = \frac{\tilde{P}_t}{P_t} = 1 + r_t.$  
and the Binary (upward vs.\ downward movement) as $\text{Binary}_t = \mathbb{1}[\tilde{P}_t > P_t] =
        \begin{cases}
            1, & \tilde{P}_t > P_t, \\
            0, & \tilde{P}_t \leq P_t.
        \end{cases}$

\paragraph{Metrics:}
Predictive performance is quantified 
using 
classification and regression metrics, depending on the target formulation. 
For instance, to evaluate the efficiency in predicting directional (up or down) stock trends, we compute the F1-score, and the Area Under the ROC Curve (AUC).  
For regression tasks, i.e. predicting the stock price change, we compute the Root Mean Squared Error (RMSE) and Mean Absolute Error (MAE)
to approximate the variance proportion explained by the model.  

\paragraph{Implementation details:}
We use the LLM-based sentiment analysis model
with their default recommended setup.
In order to set a fair base of comparison and mitigate data-leakage threats, we exclude news from our dataset that has been used in training or fine-tuning these models.
Ensemble model classifiers are trained on 80\% of SentFin dataset (80-20 Training-test split).

The stock market datasets are temporally split into subsets of 70\% for training, 10\% for validation, and 20\% for test to avoid look-ahead bias and ensure that model evaluation reflects real-world forecasting conditions.
The data is structured into fixed-length sequences (rolling window) of 30 consecutive days for training, enabling the models to capture temporal dependencies over approximately one month of trading activity. Prior to training, all numerical features are normalised (Min-Max).
Distinct stock prediction models have been trained (10 times each with different random seeds), offering all possible combinations of considered forecasting architectures for every target stock, for every prediction task, using or not using sentiment features from every sentiment analysis model. 

We conduct the experiments on a standard computing environment\footnote{NVIDIA GeForce GTX 1650 Ti GPU with 4 GB VRAM, 
a 2.60 GHz CPU and 16 GB of RAM. The system is equipped with CUDA 12.7 and driver version 565.77.01.}.
We implement the experimental study pipelines in Python
using well-known open-source data manipulation and machine learning libraries.
We make our complete codebase and setup instructions with additional details available~\footnote{\url{https://github.com/Walids35/capstone-stock-prediction.git}}
to enable reproducibility and support similar future research.

\section{Results and discussion}
\label{sec:results}

\subsection{Effectiveness and complementarity of sentiment analysis models}
\label{subsec:rq1}

Although being a generic-purpose LLM, DeBERTa outperforms both RoBERTa and FinBERT, predicting sentiments accurately for over 75\% of the dataset, followed by FinBERT then RoBERTa scoring both below 70\%; 70.9\% and 58.9\%, respectively.

\begin{figure*}[t]
\centering
\vspace{-12pt}
\begin{subfigure}[t]{0.32\textwidth}
  \centering
  \adjincludegraphics[width=\textwidth, trim={{.078\width} {.07\width} {0.078\width} {.046\width}} ,clip]{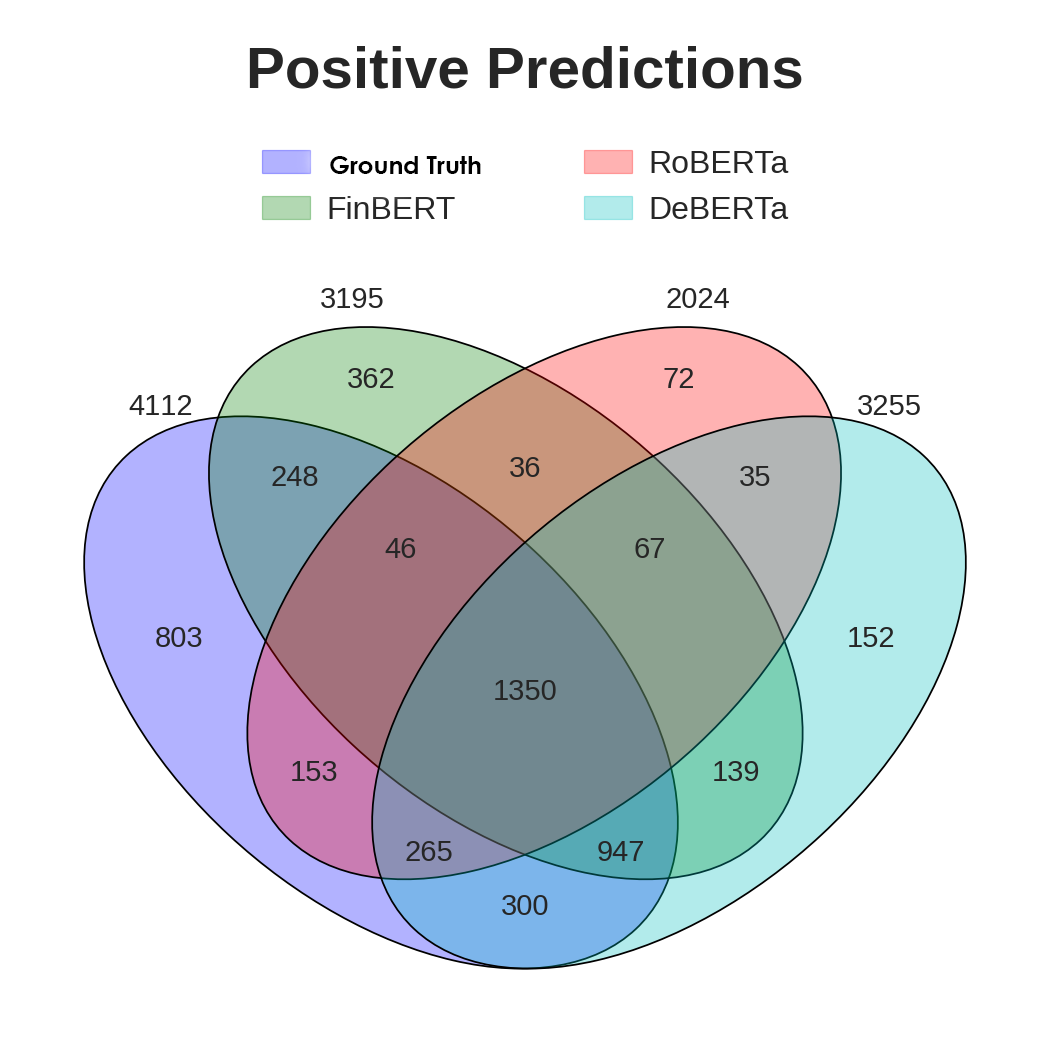}
\end{subfigure}\hfill
\begin{subfigure}[t]{0.32\textwidth}
  \centering
  \adjincludegraphics[width=\textwidth, trim={{.078\width} {.07\width} {0.078\width} {.046\width}} ,clip]{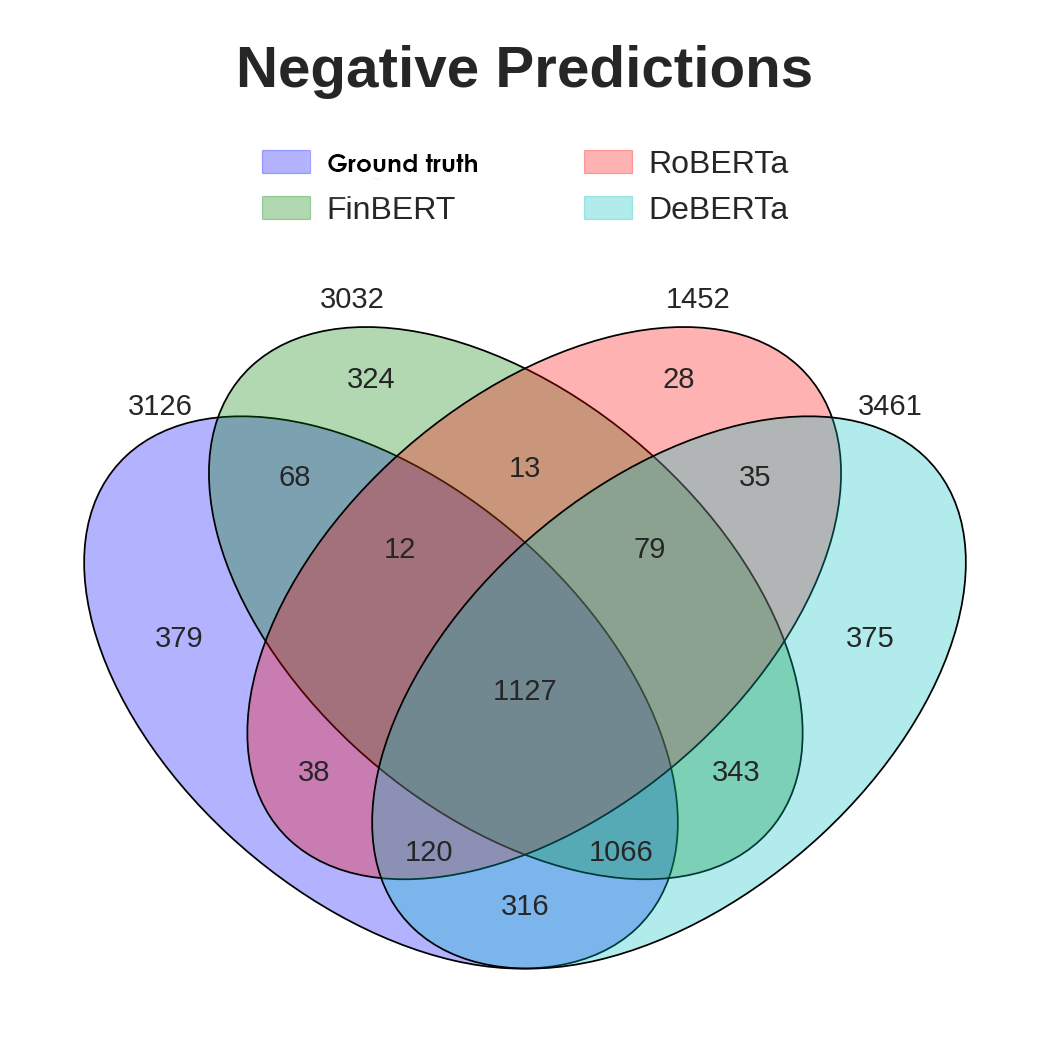}
\end{subfigure}\hfill
\begin{subfigure}[t]{0.32\textwidth}
  \centering
  \adjincludegraphics[width=\textwidth, trim={{.078\width} {.07\width} {0.078\width} {.046\width}} ,clip]{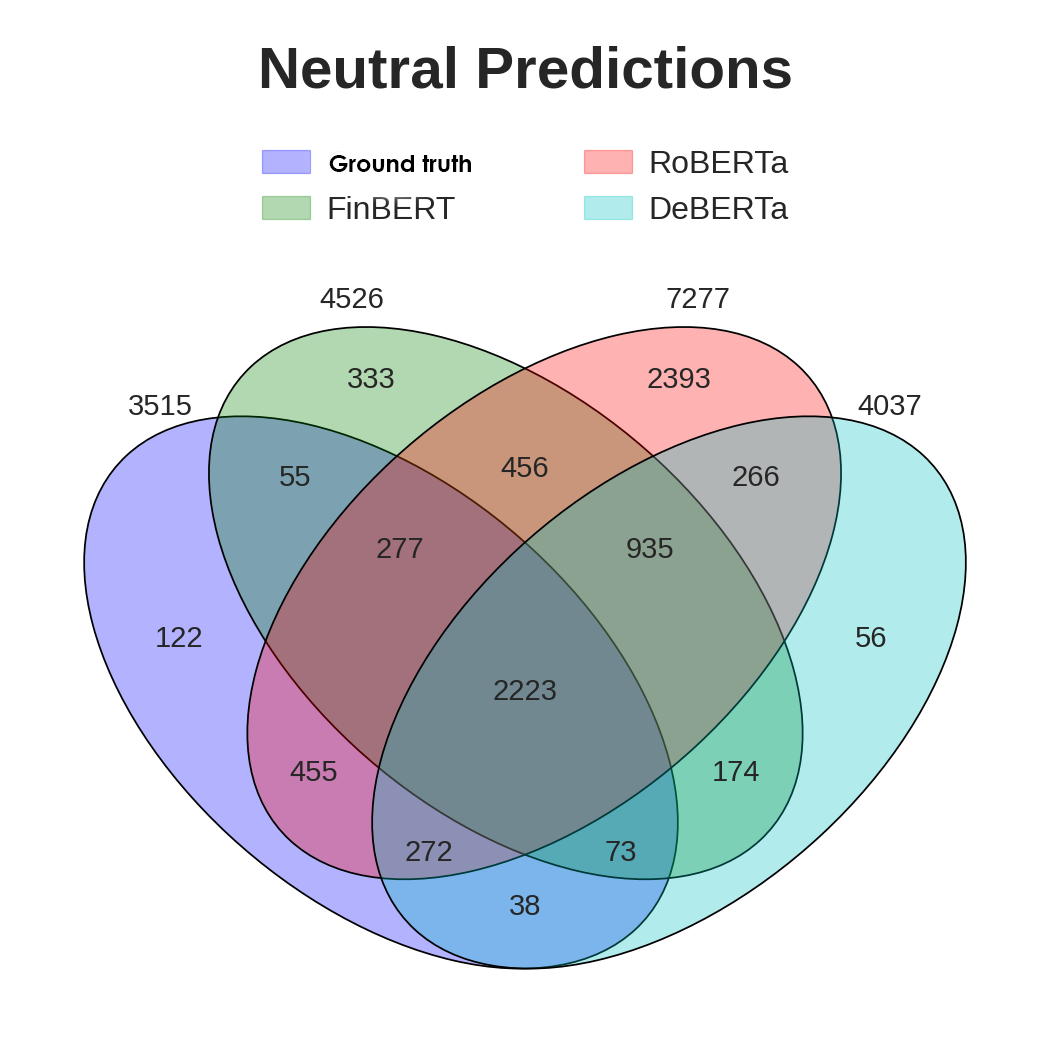}
\end{subfigure}
\vspace{-8pt}
\caption{Venn diagrams showing overlap among Original labels, FinBERT, RoBERTa, and DeBERTa across sentiment classes.}
\label{fig:venn_pos} 
\end{figure*}

\begin{minipage}{0.47\linewidth}
The overlap and disjoint sets of accurately predicted news presented in the Venn diagrams of Figure~\ref{fig:venn_pos} endorse the complementarity between the models, showing that they do not accurately predict the same news of the same classes. 
This motivates the training of a learning approach which would learn how to combine results from each model in order to improve the sentiment prediction performance.

\end{minipage}\hfill
\begin{minipage}{0.51\linewidth}
\centering
\captionof{table}{\small Performance of ensemble classifiers trained on combined outputs from FinBERT, RoBERTa, and DeBERTa.}
\vspace{-10pt}
\label{tab:ensemble_results}
\begin{center}
\scalebox{0.70}{
\begin{tabular}{lcccc}
\toprule
\textbf{Model} & \textbf{Accuracy} & \textbf{Precision} & \textbf{Recall} & \textbf{F1-Score} \\
\midrule
FinBERT & 0.696 & 0.713 & 0.701 & 0.7 \\
RoBERTa & 0.589 & 0.744 & 0.591 & 0.585 \\
DeBERTa & 0.752 & 0.761 & 0.759 & 0.755 \\
Logistic Regression & 0.778 & 0.791 & 0.778 & 0.782 \\
Random Forest       & 0.788 & 0.796 & 0.788 & 0.790 \\
SVM                 & \textbf{0.791} & \textbf{0.799} & \textbf{0.791} & \textbf{0.793} \\
\bottomrule
\end{tabular}}
\end{center}
\end{minipage}

The results reported in Table~\ref{tab:ensemble_results} endorse this idea, showing an advantage in employing an ensemble model over using any LLM separately, i.e. an SVM ensemble model achieves on average about 80\% of accuracy, precision, recall and F1 score.

\subsection{Comparison of different daily news sentiment aggregation methods}

To assess the contribution of sentiment representations (aggregation of daily news), we analyze the performance of LSTM when different sentiment features are incorporated into market data. Specifically, we design four ablation experiments to isolate the effect of each sentiment type: LSTM (using all sentiment features), LSTM{\_}wo{\_}count (excluding news count per label), LSTM{\_}wo{\_}sum (excluding sentiment score aggregation), LSTM{\_}wo{\_}count{\_}sum (excluding both news count and sentiment aggregation), and LSTM{\_}wo{\_}majority (excluding majority vote aggregation). The performance of these variants is then evaluated across all output types.

For the binary price prediction task (Table ~\ref{tab:binary_price}, the baseline LSTM model achieves the best overall balance across metrics, with an AUC of 0.5557, Accuracy of 0.5747, F1-Score of 0.5507, Precision of 0.6042, and Recall of 0.5747. The ablation variants generally result in reduced performance, confirming the importance of each component. Specifically, removing the count feature (LSTM{\_}wo{\_}count) lowers both AUC (0.5393) and F1-Score (0.5361), while the sum feature also contributes positively, as its removal (LSTM{\_}wo{\_}sum) decreases AUC to 0.5468 and F1-Score to 0.5411.

\begin{minipage}{0.647\linewidth}

The combination removal (LSTM{\_}wo{\_}count{\_}sum) produces the weakest outcome with an AUC of 0.5318 and F1-Score of 0.5251, showing the compounding effect of feature reduction. Interestingly, excluding the majority component has minimal impact, with LSTM{\_}wo{\_}majority maintaining similar performance to the baseline. These results highlight that while the baseline LSTM provides the strongest predictive power, count and sum features play a key role in supporting accurate classification.

\end{minipage}\hfill
\begin{minipage}{0.335\linewidth}
\captionof{table}{\small Binary price prediction performance for LSTM and its ablation variants.}
\vspace{-10pt}
\label{tab:ensemble_results}
\begin{center}
\scalebox{0.70}{
\begin{tabular}{lccccc}
\toprule
\textbf{Variant}            & \textbf{AUC}     & \textbf{F1-Score}  \\
\midrule
LSTM               & 0.5557    & 0.5507    \\
LSTM\_wo\_count     & 0.5393    & 0.5361    \\
LSTM\_wo\_count\_sum & 0.5318    & 0.5251   \\
LSTM\_wo\_sum       & 0.5468    & 0.5411  \\
LSTM\_wo\_majority  & 0.5473    & 0.5495  \\
\bottomrule
\end{tabular}}
\label{tab:binary_price}
\end{center}
\end{minipage}

\subsection{Effectiveness of sentiment analysis models on stock movement classification}

\ak{4 arch as columns : NS(F1(avg,sd), AUC (avg,sd)), 6 sentiment model (F1(avg,sd), AUC(avg,sd)), }

\begin{table}[h]

\centering
\caption{\small Performance comparison of sentiment analysis models on stock movement trend prediction (binary classification: stock price going up or down). NS stands for "No Sentiment" and refers to a baseline stock prediction without any news sentiment feature.}
\vspace{-8pt}
\resizebox{\textwidth}{!}{
\begin{tabular}{lcccccccc}
\toprule
& \multicolumn{2}{c}{LSTM} 
& \multicolumn{2}{c}{PatchTST} 
& \multicolumn{2}{c}{TimesNET} 
& \multicolumn{2}{c}{tPatchGNN} \\
\cmidrule(lr){2-3} \cmidrule(lr){4-5} \cmidrule(lr){6-7} \cmidrule(lr){8-9}
Model 
& F1 (avg, sd) & AUC (avg, sd) 
& F1 (avg, sd) & AUC (avg, sd) 
& F1 (avg, sd) & AUC (avg, sd) 
& F1 (avg, sd) & AUC (avg, sd) \\
\midrule

NS 
& 0.541 ± 0.078 & 0.545 ± 0.066 
& 0.519 ± 0.074 & 0.511 ± 0.048
& 0.534 ± 0.087 & 0.521 ± 0.077 	
& 0.430 ± 0.114 & 0.466 ± 0.060 \\

FinBERT
& \textbf{0.564 ± 0.057} &	\textbf{0.562 ± 0.055}
& 0.507 ± 0.074 &	0.487 ± 0.052 
& \textbf{0.544 ± 0.065} & 0.529 ± 0.053 
& 0.431 ± 0.112 & 0.491 ± 0.049 \\

DeBERTa
& 0.534 ± 0.075& 0.547 ± 0.038
& 0.485 ± 0.079& 0.491 ± 0.057	
& 0.540 ± 0.068 & \textbf{0.533 ± 0.057}
& 0.444 ± 0.108 & 0.498 ± 0.052 \\

RoBERTa
& 0.561 ± 0.039 & 0.564 ± 0.037
& 0.502 ± 0.087 & 0.495 ± 0.056
& 0.538 ± 0.064 & 0.514 ± 0.053	
& 0.443 ± 0.108 & 0.478 ± 0.061\\

LR
& 0.552 ± 0.062 &	0.559 ± 0.046	
& 0.511 ± 0.081 & 0.507 ± 0.057	
& 0.536 ± 0.057 & 0.522 ± 0.054	
& \textbf{0.461 ± 0.113} & \textbf{0.500 ± 0.063} \\

RF
& 0.539 ± 0.054&	0.549 ± 0.037	
& 0.508 ± 0.070	&0.505 ± 0.054	
& 0.509 ± 0.074	&0.506 ± 0.049	
& 0.449 ± 0.114&	0.495 ± 0.062 \\

SVM
& 0.554 ± 0.049	&0.553 ± 0.041	
&	\textbf{0.519 ± 0.076} &	\textbf{0.514 ± 0.057}
&	0.528 ± 0.061&	0.510 ± 0.050	
&	0.411 ± 0.109&	0.487 ± 0.054\\

\bottomrule
\end{tabular}
\label{tab:trend}
}
\end{table}
Table~\ref{tab:trend} reports the performance of the different stock prediction models (columns) when fed with daily news predicted-sentiment features (rows), in terms of average F1-score, AUC and their corresponding standard deviation computed accross 10 different runs. 


The results depict also that LSTM and TimesNET obtain higher performance values compared to PatchTST and tPatchGNN across most configurations. Under the LSTM architecture, the highest F1-score is achieved by SVM (0.554 ± 0.049), while the highest AUC is obtained by FinBERT (0.562 ± 0.055). For PatchTST, SVM records the best performance in both F1 (0.519 ± 0.076) and AUC (0.514 ± 0.057). Within TimesNET, DeBERTa achieves the highest F1 (0.540 ± 0.068) and AUC (0.533 ± 0.057). In contrast, tPatchGNN shows lower values overall, with Logistic Regression obtaining the highest F1 (0.461 ± 0.113) and AUC (0.500 ± 0.063) within that architecture.

When comparing sentiment-based models (FinBERT, DeBERTa, RoBERTa) to the no-sentiment (NS) baseline, increases in AUC are observed for LSTM with FinBERT (0.562 vs. 0.521) and for TimesNET with DeBERTa (0.533 vs. 0.521). In other cases, the differences between sentiment-based models and the NS baseline remain relatively small across architectures.

This indicates, that although SVM stacking approach predicts more accurately sentiments in the FPB dataset (Table~\ref{tab:ensemble_results}), it does not always provide the best guidance fo stock movement prediction for every approach. 
Moreover, models do not benefit equally from the added news sentiment features.  
Interestingly, we notice that LSTM and PatchTST are more suited for the classification task than TimesNET and tPatchGNN.
\subsection{Effectiveness of sentiment analysis models on stock movement regression}
\ak{price factor: 4 arch as columns : NS(MAE(avg,sd), RSE (avg,sd)), 6 sentiment model (MAE(avg,sd), RSE(avg,sd))}

\begin{table}[h]
\centering
\caption{\small Performance comparison of sentiment analysis models on stock price movement prediction; regression task on the price factor. NS stands for "No Sentiment" and refers to a baseline stock prediction without any news sentiment feature.}
\vspace{-8pt}
\resizebox{\textwidth}{!}{
\begin{tabular}{lcccccccc}
\toprule
& \multicolumn{2}{c}{LSTM} 
& \multicolumn{2}{c}{PatchTST} 
& \multicolumn{2}{c}{TimesNET} 
& \multicolumn{2}{c}{tPatchGNN} \\
\cmidrule(lr){2-3} \cmidrule(lr){4-5} \cmidrule(lr){6-7} \cmidrule(lr){8-9}
Model 
& MAE (avg, sd) & RSE (avg, sd) 
& MAE (avg, sd) & RSE (avg, sd) 
& MAE (avg, sd) & RSE (avg, sd) 
& MAE (avg, sd) & RSE (avg, sd) \\
\midrule

NS & \textbf{ 0.032 ± 0.009}&\textbf{2.070 ± 0.663}&0.391 ± 0.186&5.030 ± 2.138&0.483 ± 0.215&6.399 ± 2.593&\textbf{0.170 ± 0.085}&\textbf{2.188 ± 0.750} \\

FinBERT&0.033 ± 0.010&2.116 ± 0.694&0.251 ± 0.166&3.493 ± 2.293&0.274 ± 0.145&3.769 ± 1.794&0.176 ± 0.088&2.247 ± 0.781 \\

RoBERTa&0.034 ± 0.011&2.180 ± 0.727&0.237 ± 0.149&3.384 ± 2.064&\textbf{0.254 ± 0.158}&\textbf{3.620 ± 2.197}&0.176 ± 0.087&2.244 ± 0.779\\

DeBERTa&0.033 ± 0.010&2.098 ± 0.671&0.208 ± 0.148&3.020 ± 2.113&0.291 ± 0.146&3.967 ± 1.840&0.175 ± 0.087&2.239 ± 0.777\\

LR&0.033 ± 0.010&2.112 ± 0.689&0.218 ± 0.133&3.038 ± 1.928&0.298 ± 0.157&3.914 ± 1.926&0.178 ± 0.088&2.270 ± 0.791\\
RF&0.034 ± 0.010&2.136 ± 0.705&0.217 ± 0.125&3.078 ± 2.152&0.337 ± 0.179&4.332 ± 2.208&0.179 ± 0.088&2.279 ± 0.794\\
SVM&0.033 ± 0.010&2.103 ± 0.678&\textbf{0.205 ± 0.122}&\textbf{2.881 ± 1.580}&0.306 ± 0.165&4.016 ± 2.076&0.178 ± 0.088&2.271 ± 0.791\\

\bottomrule
\end{tabular}
\label{tab:factor}
}
\end{table}

Table~\ref{tab:factor} reports the performance of the different stock prediction models (columns) when fed with daily news predicted-sentiment features (rows), in terms of average MAE, RSE and their corresponding standard deviation computed accross 10 different runs. Lower error values imply better performance. 

The influence of sentiment features on forecasting accuracy varies noriceably across architectures. LSTM and tPatchGNN exhibit minimal sensitivity to sentiment integration, with differences between the NS baseline and sentiment-augmented models remaining within 0.002-0.009 for MAE and 0.051-0.110 for RSE, and both architectures achieving their lowest errors without sentiment features. In contrast, PatchTST and TimesNET demonstrate substantial performance gains from sentiment incorporation. For PatchTST, sentiment models reduce MAE by 0.140-0.186 and RSE by 1.537-2.149 compared to the NS baseline (0.391 MAE, 5.030 RSE), with SVM-based sentiment yielding the largest improvement (0.205 MAE, 2.881 RSE). 

TimesNET shows even more pronounced benefits, where sentiment integration reduces MAE by 0.146-0.229 and RSE by 2.067-2.779 relative to NS (0.483 MAE, 6.399 RSE), with RoBERTa producing the strongest gains (0.254 MAE, 3.620 RSE). Across sentiment modeling approaches, traditional ML methods (SVM, LR) and transformer-based models (DeBERTa, RoBERTa, FinBERT) show comparable effectiveness within each architecture, with no single sentiment extraction method consistently outperforming others across all forecasting models.

\section{Conclusion}
\label{sec:conclusion}

In this work, we investigate the financial news sentiment analysis performance in practice, particularly its impact on stock price movement prediction. 
We train and compare the performance of SOTA fundamentally different stock price movement prediction architectures, with different sentiment analysis models and different daily news sentiment representations. 
The results show that the studied sentiment analysis models perform efficiently, with DeBERTa outperforming both FinBERT and RoBERTa. 
Moreover, combining the three models, i.e. via an SVM algorithm, offers more accurate predictions (about 80\%).
When introduced in the stock price prediction tasks, the sentiment information improved the regression performance of models based on PatchTST and TimesNet, as well as the accuracy and AUC of LSTM-, PatchTST and tPatchGNN-based classifiers. 

\section*{Acknowledgment}
This research was funded in whole, or in part, by the Luxembourg National Research Fund (FNR), grant reference NCER22/IS/16570468/NCER-FT.

\bibliography{iclr2026_conference}
\bibliographystyle{iclr2026_conference}

\end{document}